# Multidimensional representations in late-life depression: convergence in neuroimaging, cognition, clinical symptomatology and genetics

Junhao Wen, PhD[1,*], Cynthia H.Y. Fu, MD, PhD [2,3], Duygu Tosun, PhD [4], Yogasudha Veturi, PhD [5], Zhijian Yang, MS[1], Ahmed Abdulkadir, PhD [1], Elizabeth Mamourian, MS[1], Dhivya Srinivasan, MS[1], Jingxuan Bao, MS[6], Guray Erus, PhD [1], Haochang Shou, PhD [1,7], Mohamad Habes, PhD [8], Jimit Doshi, MS[1], Erdem Varol, PhD [9], Scott R Mackin, PhD [10], Aristeidis Sotiras, PhD [11], Yong Fan, PhD [1], Andrew J. Saykin, PhD [12], Yvette I. Sheline, MD, PhD [13], Li Shen, PhD [6], Marilyn D. Ritchie, PhD [5], David A. Wolk, MD, PhD [1,14], Marilyn Albert, PhD [15], Susan M. Resnick, PhD [16], Christos Davatzikos, PhD [1,*,&]

[1]Center for Biomedical Image Computing and Analytics, Perelman School of Medicine, University of Pennsylvania, Philadelphia, USA
[2] University of East London, School of Psychology, London, UK
[3] Centre for Affective Disorders, Institute of Psychiatry, Psychology and Neuroscience, King's College London, London, UK
[4]Department of Radiology and Biomedical Imaging, University of California, San Francisco, CA, USA
[5]Department of Genetics and Institute for Biomedical Informatics, Perelman School of Medicine, University of Pennsylvania, Philadelphia, PA, USA
[6]Department of Biostatistics, Epidemiology and Informatics University of Pennsylvania Perelman School of Medicine, Philadelphia, PA 19104, USA
[7]Penn Statistics in Imaging and Visualization Center, Department of Biostatistics, Epidemiology, and Informatics, Perelman School of Medicine, University of Pennsylvania, Philadelphia, USA
[8]Glenn Biggs Institute for Alzheimer's & Neurodegenerative Diseases, University of Texas Health Science Center at San Antonio, San Antonio, USA
[9]Department of Statistics, Center for Theoretical Neuroscience, Zuckerman Institute, Columbia University, New York, USA
[10]Department of Psychiatry, University of California, San Francisco, CA, USA
[11]Department of Radiology and Institute for Informatics, Washington University School of Medicine, St. Louis, USA
[12]Radiology and Imaging Sciences, Center for Neuroimaging, Department of Radiology and Imaging Sciences, Indiana Alzheimer's Disease Research Center and the Melvin and Bren Simon Cancer Center, Indiana University School of Medicine, Indianapolis
[13]Center for Neuromodulation in Depression and Stress, Department of Psychiatry, University of Pennsylvania Perelman School of Medicine, Philadelphia, Pennsylvania, USA
[14]Department of Neurology and Penn Memory Center, University of Pennsylvania, Philadelphia, USA
[15]Department of Neurology, Johns Hopkins University School of Medicine, USA
[16]Laboratory of Behavioral Neuroscience, National Institute on Aging, Baltimore, USA
[&] For the iSTAGING consortium, the ADNI, the BIOCARD, and the BLSA studies

[*]Corresponding authors:
Junhao Wen, PhD – junhao.wen89@gmail.com
Christos Davatzikos, PhD – Christos. Davatzikos@pennmedicine.upenn.edu
3700 Hamilton Walk, Philadelphia, PA 19104

**Search terms:** late-life depression; heterogeneity; semi-supervised clustering; dimensional representation
**Word counts**: 2995

# Key points

**Question:** Is late-life depression (LLD) associated with one or multiple structural neuroimaging patterns?

**Findings**: Two dimensions best represented LLD neuroanatomical heterogeneity. Dimension 1 was associated with preserved brain structure, whereas Dimension 2 demonstrated diffuse structural abnormalities and greater cognitive impairment. One *de novo* independent genetic variant was significantly associated with Dim1 but not with Dim2. Notably, the two dimensions manifested significant genetic heritability in the general population, and Dim2 was longitudinally more vulnerable to Alzheimer's disease and brain aging than Dim1.

**Meanings**: The two dimensions encompass heterogeneity in LLD and offer the potential for clinical precision in diagnosis and prognosis.

## ABSTRACT

**Importance:** Late-life depression (LLD) is characterized by considerable heterogeneity in clinical manifestation. Unraveling such heterogeneity would aid in elucidating etiological mechanisms and pave the road to precision and individualized medicine.

**Objective:** We sought to delineate, cross-sectionally and longitudinally, disease-related heterogeneity in LLD linked to neuroanatomy, cognitive functioning, clinical symptomatology, and genetic profiles.

**Design & setting:** The iSTAGING study is an international multicenter consortium investigating brain aging in pooled and harmonized data from 13 studies with over 35,000 participants, including a subset of individuals with major depressive disorders.

**Participants:** Multimodal data from a multicentre sample ($N$=996), including neuroimaging, neurocognitive assessments, and genetics: 501 LLD participants (332 women, mean age 67.39 $\pm$ 5.56 years) and 495 healthy controls (333 women, mean age 66.53 $\pm$ 5.16 years) were analyzed. A semi-supervised clustering method (HYDRA) was applied to regional grey matter (GM) brain volumes to derive dimensional representations.

**Exposure**: None

**Main outcome and Measure**: Two dimensions were identified, which accounted for the LLD-related heterogeneity in voxel-wise GM maps, white matter (WM) fractional anisotropy (FA), neurocognitive functioning, clinical phenotype, and genetics.

**Results:** Dimension one (Dim1) demonstrated relatively preserved brain anatomy without WM disruptions relative to healthy controls. In contrast, dimension two (Dim2) showed widespread brain atrophy and WM integrity disruptions, along with cognitive impairment and higher depression severity. Moreover, one *de novo* independent genetic variant (rs13120336) was significantly associated with Dim 1 but not with Dim 2. Notably, the two dimensions demonstrated significant SNP-based heritability of 18-27% within the general population ($N$=12,518 in UKBB). Lastly, in a subset of individuals having longitudinal measurements, Dim2 demonstrated a more rapid longitudinal change in GM and brain age, and was more likely to progress to Alzheimer's disease, compared to Dim1 ($N$=1,413 participants and 7,225 scans from ADNI, BLSA, and BIOCARD datasets).

**Conclusions and Relevance:** Heterogeneity in LLD was represented by two dimensions with distinct neuroanatomical, cognitive, clinical, and genetic profiles. This dimensional approach provides a novel mechanism for investigating the heterogeneity of LLD and the relevance of the latent dimensions to possible disease mechanisms, clinical outcomes, and responses to interventions.

## Introduction

Major depressive disorder (MDD) is one of the most common mental health disorders and is a leading contributor to disability worldwide (1, 2). Late-life depression (LLD) generally refers to MDD, late-onset or early-onset, from 60-65 years of age, which affects up to 18% of older adults in the general community but over 30% of those in care homes (3–6).

There is considerable heterogeneity in the presentation and progression of clinical symptomatology, comorbid psychiatric, medical, neurological disorders, and course of illness (7, 8). Pharmacological and psychological treatments tend to be less effective in LLD. Up to 50% of LLD patients do not achieve remission with their first treatment (9, 10). LLD is associated with high comorbidity, including cardiac and cerebrovascular disease (11), stroke (12), as well as increased risk for obesity, diabetes, frailty (13), and neurodegenerative diseases such as Alzheimer's disease and vascular dementia (14–17). Several hypotheses for underlying neuropathological mechanisms have been proposed to account for its high heterogeneity: depression-executive dysfunction syndrome, vascular depression, and inflammation hypothesis (7).

T1-weighted magnetic resonance imaging (MRI) has revealed grey matter (GM) reductions in bilateral anterior cingulate and medial frontal cortices, insula, putamen, and globus pallidus, extending into the parahippocampal gyrus, amygdala, and hippocampus, while larger GM volumes have been observed in the lingual gyrus (18), putamen and caudate regions (19). Diffusion tensor imaging (DTI) demonstrates widespread losses in white matter (WM) integrity, including in the anterior thalamic radiation, cingulum, corticospinal tract, superior and inferior longitudinal fasciculi, and uncinate fasciculus (20). Collectively, the findings support biological models of LLD being associated with cortical atrophy and white matter abnormalities in specific brain networks, although the extent and magnitude of neuroimaging findings have varied across cohort studies.

A developing body of methodological advancement in data-driven biological subtypes (21–26) is challenging the traditional definition of neurological diseases, such as Alzheimer's disease (21, 22, 24, 25). It has captured increasing attention and advocated that distinct neuropathological mechanisms may underlie heterogeneity in the presentation and progression of the clinical phenotype. Furthermore, the extent to which genetic heterogeneity influences or interacts with the phenotypic expression was barely explored (27). Individual-level variability, including environment, genetic or other factors, may lead to different levels of genetic liability to the disease (28).

We sought to delineate the heterogeneity in LLD in a large multicenter sample ($N = 996$) using a state-of-the-art semi-supervised clustering method (HYDRA) (29). We hypothesized that multiple distinct dimensions coexist to account for the underlying heterogeneity and that these dimensions might be prominent in the general population and longitudinal trajectories.

## Materials and methods

### Participants

The iSTAGING study is an international consortium consisting of various imaging protocols, scanners, data modalities, and pathologies (30). It currently comprises harmonized MRI data from more than 35,000 participants encompassing a wide range of ages (22 - 90 years) from more than 13 studies. The present study includes LLD patients from four cohorts, including the UK Biobank (UKBB) (31), Psychotherapy Response Study at the University of California San Francisco (UCSF), Baltimore Longitudinal Study of Aging (BLSA) (32, 33), and Biomarkers of Cognitive Decline Among Normal Individuals at the Johns Hopkins University (BIOCARD).

We applied a harmonized LLD definition criterion to consolidate LLD participants from the four sites and excluded participants with comorbid medical and neurological diseases that were potential confounds. Details of datasets, participant inclusion/exclusion criteria, image protocols, modality, demographics, clinical scores, and acquisition parameters for all sites are presented in **Supplementary eMethod 1** and **Supplementary eTable 1**. A total of 996 participants (501 LLD patients and 495 healthy control subjects) were included in the current study.

**Image preprocessing**

The quality-controlled (QC) (**Supplementary eMethod 2**) images were corrected for magnetic field intensity inhomogeneity (34). A state-of-the-art multi-atlas parcellation method (MUSE) (35) was used to extract regions of interest (ROI) values of the segmented GM tissue maps (**Supplementary eTable 2**). Voxel-wise regional volumetric maps (RAVENS) for each tissue volume (36) were generated by spatially aligning the skull-stripped images to a template residing in the MNI-space using a registration method (37). Fractional anisotropy (FA) maps were used to examine microstructural integrity disruptions in WM (**Supplementary eMethod 3**). The mean FA values were extracted within the 48 WM tracts of the JHU ICBM-DTI-81 WM label atlas (38). The inter-site image harmonization of the GM MUSE ROIs is detailed in **Supplementary eMethod 4.**

**Genetic preprocessing**

We consolidated an imaging-genetic dataset from UKBB that passed the QC protocol, resulting in 20,438 participants and 8,430,655 single nucleotide polymorphisms (SNPs) (**Supplementary eMethod 8**). We then selected 774 UKBB participants that overlapped with the LLD population for genetic analyses.

**Discovery of the multidimensional representation via HYDRA**

We applied a semi-supervised clustering method, termed HYDRA (29) (**Supplementary eMethod 5**), to the harmonized MUSE ROIs. Briefly, HYDRA aims to cluster disease effects instead of directly clustering patients by comparing the patterns between healthy controls (CN) and LDD patients, thus resulting in a "*1-to-k*" mapping (*k*, number of dimensions/clusters) from the CN to the patient domain. One of the advantages of HYDRA and semi-supervised clustering is that it tends to avoid clustering patients according to disease-irrelevant confounds by directly clustering differences between controls and patients.

In the current study, we chose the optimal number of dimensions/clusters (*k*) ranging from 2 to 8 clusters by the Adjusted Rand Index (ARI) (39). We performed additional analyses to evaluate the robustness of the optimal *k* clusters scheme. First, split-sample analyses (40) were carried out to assess whether the dimensions in each half exhibit similar neuroanatomical patterns, given that the two halves had similar cohort characteristics in terms of age, sex, and site. Secondly, we conducted leave-site-out validation (41) to examine if the dimensions are consistent across sites: i) training on UKBB only and ii) training on all sites. Lastly, a permutation test was performed to test the statistical significance with the optimal *k* cluster scheme (**Supplementary eMethod 6**).

**Evaluation of the multidimensional representation in neuroimaging, cognition, and genetics**

In order to obtain a deeper understanding of the *k* dimensions found by HYDRA, we subsequently investigated their characteristics regarding i) the GM tissue volume, ii) WM integrity, iii) cognitive functioning and depression-related variables, and iv) genetic architecture. Moreover, we investigated the expression of the *k* dimensions in the general population and longitudinal data.

### *Voxel-wise GM RAVENS regional tissue volumes*

Specifically, voxel-wise RAVENS GM maps from all sites were used to assess the differences in GM tissue volumes across. The *3dttest++* program (42) in AFNI (43) was used to detect the distinct neuroanatomical patterns of the corresponding dimensions vs. the CN group, considering age, sex, site, and ICV as covariates. Finally, for those voxels that survive the adjustment (Benjamini-Hochberg procedure), voxel-wise effect-size maps (i.e., Cohen's *f2*) were estimated for each paired comparison.

### *Regional WM integrity abnormality*

WM microstructural abnormality was assessed using the mean FA values of the 48 regional tracts from the UKBB data. Group comparisons were performed with multiple linear regression models using R (version 3.4.0, The R Foundation) (**Supplementary eMethod 9**). We include age and sex as fixed effects and group as the variable of interest. P-values were corrected, and Cohen's *f2* was computed with the same procedure as above.

### *Demographic, cognitive, and clinical variables*

Group comparisons for demographic, cognitive, and clinical variables (**Supplementary eTable 5**) were examined separately between the two dimensions. Mann–Whitney–Wilcoxon test was used for continuous variables (e.g., age) and the Chi-Square test of independence for categorical variables (e.g., sex). Moreover, a global effect size (i.e., Cohen's *d*) was also reported for continuous variables.

### *Genome-wide associations*

We performed GWAS with the derived binary dimension traits, i.e., Dim1 or Dim2 vs. CN using Plink 2[1]. FUMA online platform[2] was then used to annotate the genomic risk loci and independent significant SNPs (**Supplementary eMethod 8**).

*Evaluation of the multiple dimensions in the general population*

The trained model was applied to the external validation samples in UKBB ($N$=12,518, and age above 60 years). The dimension membership (**Fig. 3B**) and expression scores of the $k$ dimensions were derived (**Supplementary eMethod 7**).

We then examined the neuroanatomical patterns using RAVENS GM maps, demographic and cognitive functioning of the $k$ dimensions in the general population. Moreover, we calculated the genome-wide SNP-based heritability coefficient ($h^2$) using GCTA[3] (**Supplementary eMethod 8**).

*Evaluation of the multiple dimensions in longitudinal data and their progress to AD and brain aging*

The cross-sectionally trained model was applied to the longitudinal samples, diagnosed as CN at baseline, in ADNI, BLSA, and BIOCARD ($N$=1413 participants and 7225 scans, and age above 60 years). The dimension membership was derived to evaluate its longitudinal changes in MUSE GM ROIs, SPARE-AD (Spatial Patterns of Atrophy for REcognition of AD) (44), SPARE-BA (Brain Age) (45). Specifically, the Rate of Change (RC) over time in these variables for each participant was derived with a linear mixed-effects model[4] and compared across dimensions (e.g., Dim1 vs. Dim2) using a linear regression model (**Supplementary eMethod 9**).

---

[1] https://www.cog-genomics.org/plink/2.0/
[2] https://fuma.ctglab.nl/
[3] https://cnsgenomics.com/software/gcta
[4] https://www.statsmodels.org/stable/index.html

## Results

### HYDRA reveals two dimensions

The highest ARI (0.58) was achieved by a HYDRA model for $k$=2 clusters (**Supplementary eFigure 1**). The cluster assignment distribution for $k = 2$ to 8 across sites is presented in **Supplementary eTable 3**. For the optimal $k$=2 clustering scheme, out of the 501 LLD participants, 227 participants were assigned to Dimension 1 (Dim1) and 274 to Dimension 2 (Dim2). Moreover, the optimal $k$=2 clustering scheme was replicated in split-sample and leave-site-out analyses (**Supplementary eFigure1**). For the approaches of the leave-site-out analyses, the percentage overlap for the patients assigned to the same dimension was 89.12% (91.77% for UKBB, 76.41% for BLSA, 81.27% for BIOCARD, and 84.45% for UCSF). The neuroanatomical patterns of the two dimensions were similar (**Supplementary eFigure 3**) compared to the original dimension patterns (Fig. 1). For split-sample analyses, the GM patterns for the two splits were similar (**Supplementary eFigure 2**) and compared to the original dimension patterns (**Fig. 1A**). The ARI at $k$=2 was higher than the null distribution in the permutation test (P-value<0.001). Therefore, we present the results of $k$=2 for all subsequent analyses.

### Differences in GM volumetric patterns

Dim1 demonstrated greater GM tissue volume in bilateral thalamus, putamen, and caudate relative to healthy controls. Dim2 demonstrated reduced GM tissue volume in bilateral anterior and posterior cingulate gyri, superior, middle, and inferior frontal gyri, gyrus recti, insular cortices, superior, middle, and inferior temporal gyri, etc., compared to controls (**Fig. 1A**). The results of the split-sample and leave-site-out analyses supporting this result are detailed in **Supplementary eFigure 2** and **Supplementary eFigure 3**, respectively.

**Differences in WM integrity disruption**

Dim1 exhibited similar FA values compared to controls. However, Dim2 showed widespread WM disruptions, with 31 out of the 48 WM tracts demonstrating significantly lower FA values than controls but small effect sizes ($0.01 \leq$ Cohen's $f2 \leq 0.05$, **Fig. 1B**). Specifically, the middle cerebellar peduncle tract obtained the highest effect size (Cohen's $f2$=0.05). Other affected WM tracts mainly involved the frontal lobe and subcortical limbic regions (**Supplementary eTable 4**).

**Dim1 and Dim2 demonstrate differences in clinical profiles**

Dim1 showed statistically higher scores in Fluid Intelligence scores (Cohen's $d = 0.25$), but lower errors in Pairs Matching test (Cohen's $d = -0.28$) and in Patient Health Questionnaire responses (PHQ9) (Cohen's $d = -0.45$) relative to Dim2. The two dimensions did not significantly differ in age, sex, site, or other clinical variables (details in **Supplementary eTable 5**).

**Differences in genome-wide associations**

Dim1, but not Dim2, was significantly associated with one *de novo* independent variant (rs13120336 on chromosome 4) (P-value=$3.14e10^{-8}$) (**Fig. 2**). The quantile-quantile plots are presented in **Supplementary eFigure 4**.

**Expression of the two dimensions in the general population**

Applying the trained model to UKBB samples resulted in 3500 None participants (neither dimension was expressed), 2269 Dim1 participants, 3786 Dim2 participants, and 2963 Mixed individuals (both dimensions were expressed) (**Supplementary eTable 6** and **Fig. 3B**).

The neuroanatomical patterns of the two dimensions were stable (**Fig. 3A**). Dim1 showed higher scores in Fluid Intelligence scores (P-value < 1e-10, Cohen's $d$ = 0.28), but lower errors in Pairs Matching (P-value < 1e-6, Cohen's $d$ = -0.13) compared to Dim2 (**Supplementary eTable 6**). The expression scores of the two dimensions were significantly heritable in the general population. Specifically, the $h^2$ for Dim1 and Dim2 were $0.27 \pm 0.04$ (P-value<5.7e-10), and $0.18 \pm 0.04$ (P-value<1.1e-5), respectively.

**The two dimensions and longitudinal trajectories**

The aforementioned cross-sectional model yielded in 410 None participants, 301 Dim1 participants, 390 Dim2 participants, and 330 Mixed individuals in baseline images in ADNI, BLSA, and BIOCARD, which also had longitudinal follow-up data (**Supplementary eTable 7**).

The neuroanatomical patterns of the two dimensions were stable (**Fig. 4A**). The GM RC in Dim2 decreased more rapidly than in Dim1 and None groups (-0.1 < Cohen's $f2$ < 0.1), specifically in the left precentral gyrus, temporal pole, and right anterior insula (**Fig. 4B**). Moreover, the two dimensions remained independent and stable along longitudinal trajectories (**Fig. 4C**). Lastly, Dim2 showed progression of both SPARE-AD (Cohen's $f2$=0.03) and SPARE-BA (Cohen's $f2$=0.03) compared to Dim1 (**Fig. 4D**), but not at baseline.

## Discussion

Two reproducible and distinct dimensions characterized neuroanatomical heterogeneity in LLD. Dim1 showed relatively preserved brain anatomy with larger subcortical regional volumes and was associated with one *de novo* genetic variant, while Dim2 displayed widespread brain atrophy and WM integrity disruptions with impaired cognitive functioning and increased depressive severity (**Fig. 5**). Moreover, the two dimensions were manifested in the general population and were significantly heritable. Notably, Dim2 demonstrated a higher degree of progression to AD and brain aging signatures than Dim1.

The two dimensions demonstrate the extent of underlying GM heterogeneity in patients with LLD. Several imaging findings indicate overlap with prior case-control research. GM atrophy in the insula, caudate, thalamus, hippocampus regions, and frontal lobe present in Dim2, has been widely reported in previous case-control studies (46–48). Regional atrophy in the frontal lobes is observed (49, 50), which is associated with cognitive deficits as well as reports of psychotic symptoms (51). Atrophy in limbic regions involved in the regulation of emotion, behavior, and memory is also apparent. Dim2 showed brain atrophy in hippocampal regions, suggesting that Dim2 may be more involved in mood regulation due to its connections to key frontal and subcortical regions, including the amygdala, basal ganglia, and prefrontal cortex. Atrophy in the striatum has been hypothesized to be related to degeneration in the dopaminergic connections between the caudate and cortical limbic areas for mood regulation (52).

Larger brain volumes in LLD have been thought to be an effect caused by antidepressant medications (50, 53, 54). Higher right thalamus volume though was only evident in first-episode medication-naive LLD patients, and increased gray matter in the bilateral anterior cingulate cortex was found following medication wash-out (54). Similarly, patients in remission showed increased subgenual prefrontal cortex volumes relative to healthy controls (53). UKBB structural MRI data have reported that depression phenotypes were significantly associated with greater caudate and putamen volumes (19).

The two neuroanatomical dimensions identified differed significantly in microstructural integrity. Dim1 shows no significant WM abnormalities, while Dim2 consists of widespread WM abnormalities. WM lesions may play a key role in conferring vulnerability or perpetuating depressive syndromes in LLD and contributing to the observed microstructural disturbance (55). Widespread WM disruptions can persist in LLD, even excluding WM lesions from the DTI analysis (56). WM tracts connecting fronto-subcortical and fronto-limbic regions are most frequently affected, including the uncinate fasciculus (57, 58), anterior thalamic radiation, superior longitudinal fasciculus (55, 57, 59), and posterior cingulate cortex (60). Dim2 demonstrates clinical features of LLD patients that are frequently associated with more severe cognitive deterioration (61–63).

The detected genetic variant (rs13120336) was uniquely associated with Dim1. Interestingly, the two mapped genes (CCDC110 and LOC105377590) have been previously linked to cancer and diabetes (64, 65). We speculate that these genetic factors, together with the mediating effects of antidepressants, may play a key role in expressing the heterogeneity of imaging phenotype and cognitive dysfunctions in the two dimensions differently to some extent. Many studies have shown that depression is associated with different genetic variants, some of which are not replicable across or within studies (66–69). Therefore, an exact replication needs to be performed to confirm this detected variant further. In general, our dimensional approach provides clues that the genetic associations to depression might need to be revisited due to the considerable genetic heterogeneity.

The two dimensions showed significant genetic heritability of 18-27%, potentially suggesting genetic underpinnings of neuroanatomical phenotypes associated with depression in the general population. Of note, multimorbidity, such as schizophrenia or other anxiety disorders, exists in the UKBB population (70). Such comorbidities might account for the expression of the two dimensions to some extent. MDD is a common, complex trait with an estimated genetic heritability of approximately 40% (71), and the prevalence rate ranges from 7% to 13% in the general population (69). In general, our findings confirmed the high risks and lifetime prevalence of depression in the general population.

Notably, the proposed two-dimensional representation emphasizes the tremendous prognostic potential to distinguish LLD patients from LLD co-occurring or preceding other comorbidities, such as mild cognitive impairment or prodromal AD. The longitudinal results underpin this and indicate potential heterogeneity in longitudinal trajectories towards different biological processes. That is, Dim2 progressed more aggressively to an AD or brain aging signature, whereas Dim1 expressed a preserved brain anatomy. Epidemiological studies (72, 73) have consistently found that shared risk factors exist in AD and LLD, supporting depression as a prodromal feature, or a risk factor, of AD. Interestingly, the two dimensions did not longitudinally differ in cognitive impairment, in concordance with the AD pathological cascade model (74).

This work has the following limitations. First, longitudinal data in LLD are needed to confirm the added value of the proposed multidimensional representation. Additionally, replication of the GWAS findings is required when additional data are available.

## Conclusions

LLD was characterized by two dimensions linked to neuroanatomy, cognitive functioning, and genetic profiles. The two-dimensional representation offered a system for future research on the underlying etiology mechanisms, heterogeneity of genetic architectures, and the potential for individualized clinical management.

## Conflicts of interest

DAW served as Site PI for studies by Biogen, Merck, and Eli Lilly/Avid. He has received consulting fees from GE Healthcare and Neuronix. He is on the DSMB for a trial sponsored by Functional Neuromodulation. Dr. Saykin receives support from multiple NIH grants (P30 AG010133, P30 AG072976, R01 AG019771, R01 AG057739, U01 AG024904, R01 LM013463, R01 AG068193, T32 AG071444, and U01 AG068057 and U01 AG072177). He has also received support from Avid Radiopharmaceuticals, a subsidiary of Eli Lilly (in kind contribution of PET tracer precursor); Bayer Oncology (Scientific Advisory Board); Siemens Medical Solutions USA, Inc. (Dementia Advisory Board); Springer-Nature Publishing (Editorial Office Support as Editor-in-Chief, Brain Imaging and Behavior).

## Acknowledgments

This work was supported, in part, by NIH grants 1RF1-AG054409-01 and U01-AG068057. ADNI is supported by NIH grants U01-AG024904 and RC2-AG036535. The BIOCARD study is supported by NIH grant U19-AG033655. The BLSA is supported by the Intramural Research Program, National Institute on Aging, and Research and Development Contract HHSN-260-2004-00012C. This research has been conducted using the UK Biobank Resource under Application Number 35148.

## Authors' contributions:

Ph.D. Wen takes full responsibility for the integrity of the data and the accuracy of the data analysis.

*Study concept and design*: Davatzikos, Wen, Fu

*Acquisition, analysis, or interpretation of data*: Davatzikos, Wen, Fu, Tosun, Resnick

*Drafting of the manuscript*: Wen, Fu

*Critical revision of the manuscript for important intellectual content*: Wen, Fu, Tosun, Veturi, Yang, Mamourian, Srinivasan, Bao, Erus, Shou, Abdulkadir, Habes, Doshi, Varol, Mackin, Sotiras, Fan, Sheline, Saykin, Shen, Ritchie, Wolk, Albert, Resnick, Davatzikos

*Statistical and genetic analysis*: Wen

*Study supervision*: Davatzikos

**Figures**

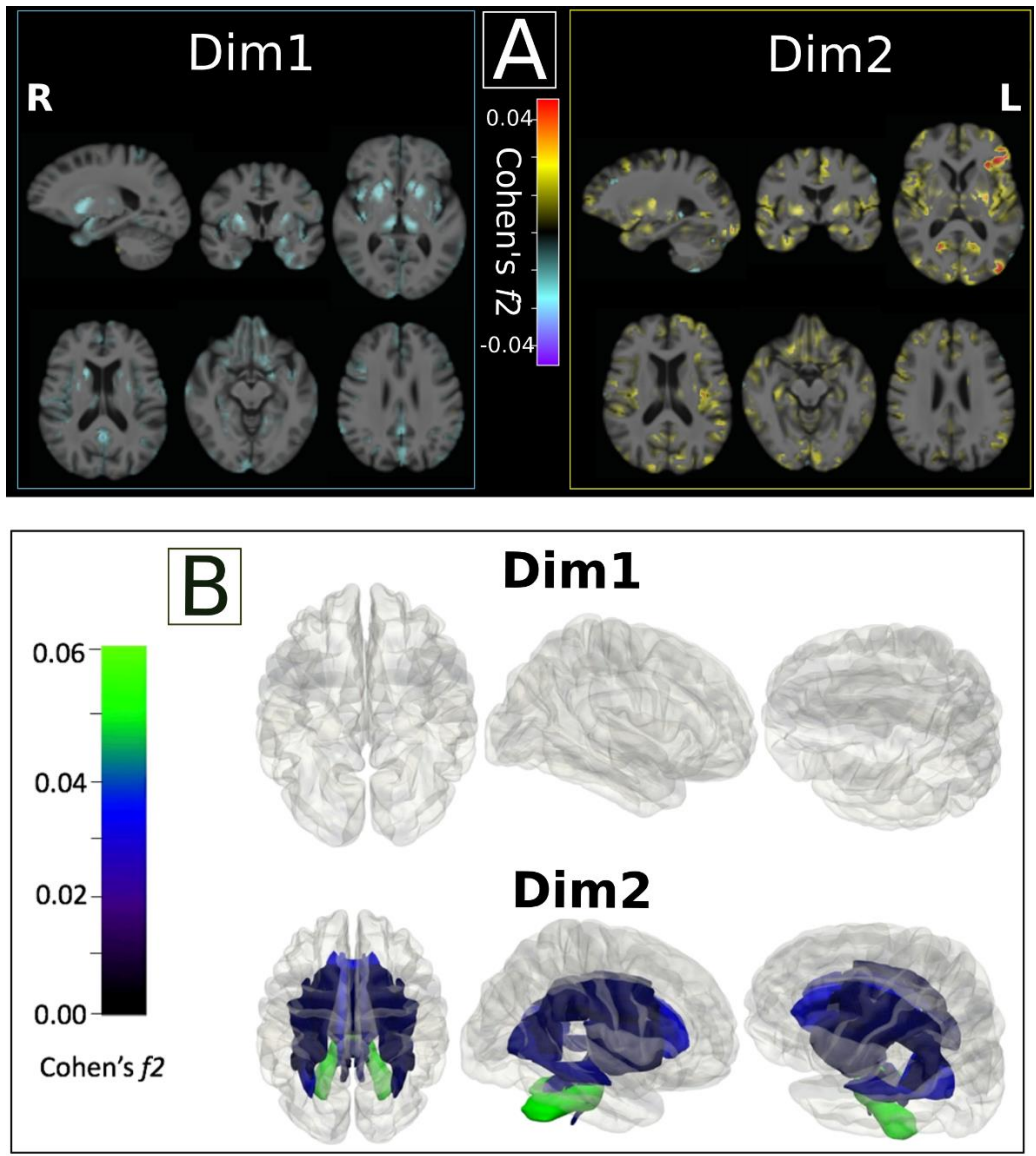

**Figure 1**: The two neuroanatomical dimensions show distinct grey matter abnormalities and white matter integrity disruptions. Effect size maps were identified in Dimension 1 (Dim1) and Dimension 2 (Dim2) compared to controls (CN), respectively. **A**) Multiple selective views are shown in different views. Warmer color denotes brain atrophy (i.e., CN > Dim), and cooler color represents larger tissue volume (i.e., Dim > CN). Both directions are shown for each dimension. L: left; R: right. The effect size map is shown in a radiological fashion, i.e., the brain's left shown to the right of the display. **B**) Dim1 and Dim2 demonstrate two distinct WM patterns based on FA values. Dim1 exhibits a normal appearance, without significant difference from controls; whereas Dim2 shows widespread disruptions in WM integrity. The P-value and effect size for all the 48 WM tracts are shown in **Supplementary eTable 4**. Both directions of the comparisons are performed, but effect sizes only show WM integrity disruptions. For references, Cohen's $f2$ of $\geq$ 0.02, $\geq$ 0.15, and $\geq$ 0.35 signify small, moderate, and large effect sizes, respectively.

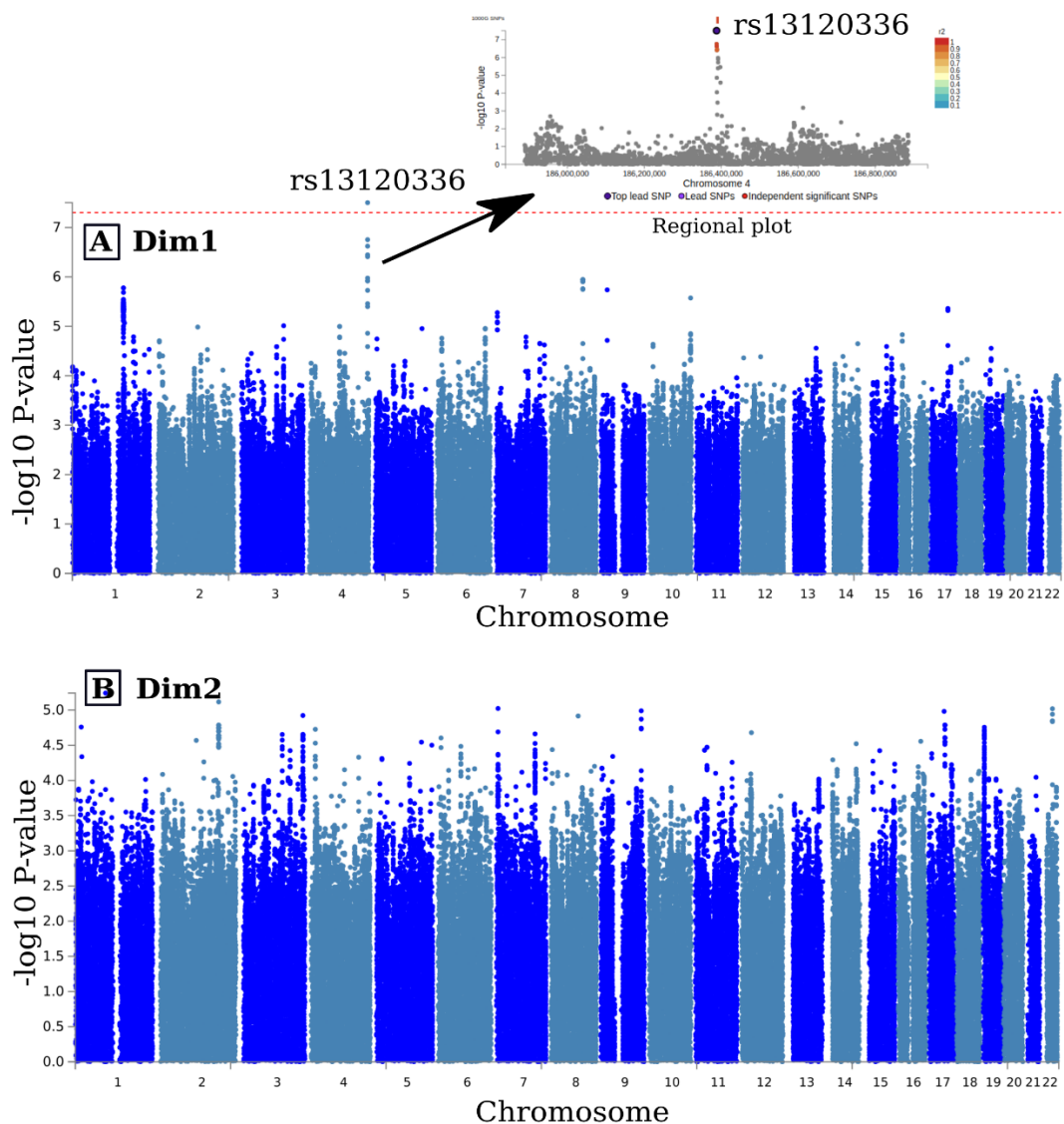

**Figure 2**: Dim1 and Dim2 demonstrate distinct profiles in GWAS. A) Dim1 was significantly associated with a novel genomic risk locus. This significant independent SNP (rs13120336) is in LD with other seven-candidate SNPs that passed the GWAS P-value threshold (5e-8). FUMA identified two corresponding protein-encoding genes: CCDC110 and LOC105377590; B) Dim2 was not significantly associated with any variants.

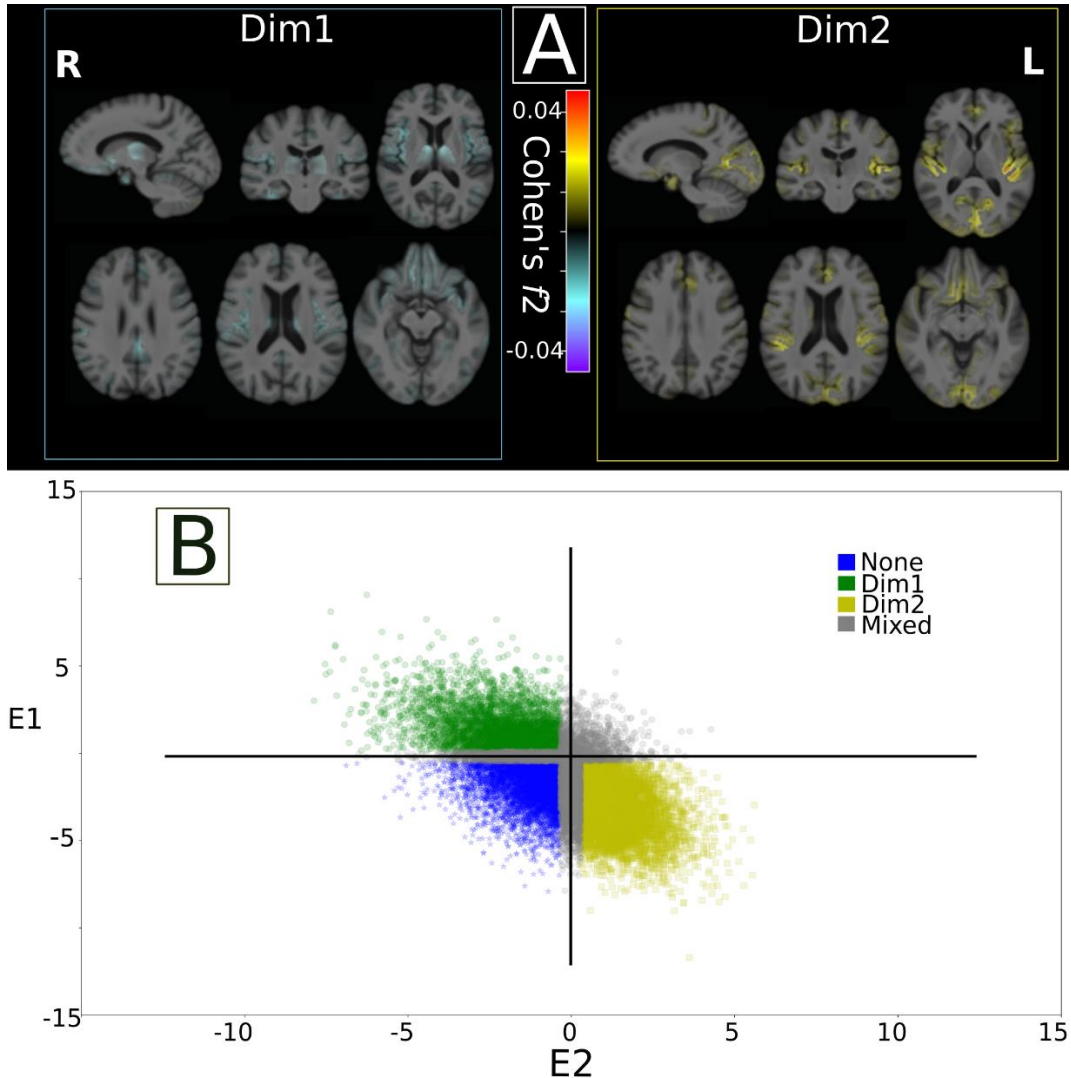

**Figure 3**: **A**) The two neuroanatomical dimensions in UKBB show distinct grey matter abnormalities. Effect size maps of GM patterns were identified in Dimension 1 (Dim1) and Dimension 2 (Dim2) compared to None (the dimension that does not express in Dim1 and Dim2), respectively. Multiple selective views are shown with the number of the slice in the axial view. Warmer color denotes brain atrophy (i.e., None > Dim), and cooler color represents larger tissue volume (i.e., Dim > None). Both directions are shown for each dimension. Cohen's $f2$ of $\geq 0.02$, $\geq 0.15$, and $\geq 0.35$ signify small, moderate, and large effect sizes, respectively. L: left; R: right. The effect size map is shown in a radiological fashion, i.e., the brain's left shown to the right of the display. We include age, sex, and ICV as fixed effects and group (None vs. Dim1 or Dim2) as the variable of interest. The likelihood ratio test was used to test each effect. **B**) The quadrant plot after applying the HYDRA model trained on the LLD population to the external UKBB individuals. X-axis and Y-axis represent the expression scores for each individual at the Dim1 and Dim2, respectively. The dimension membership was decided based on the two expression scores, E1 and E2. Specifically, the individual was assigned as None when E1 and E2 are smaller than -0.3, as Dim1 when E1 > 0.3 and E2 < -0.3, as Dim2 when E1 < -0.3 and E2 > 0.3, and as Mixed for the other individuals.

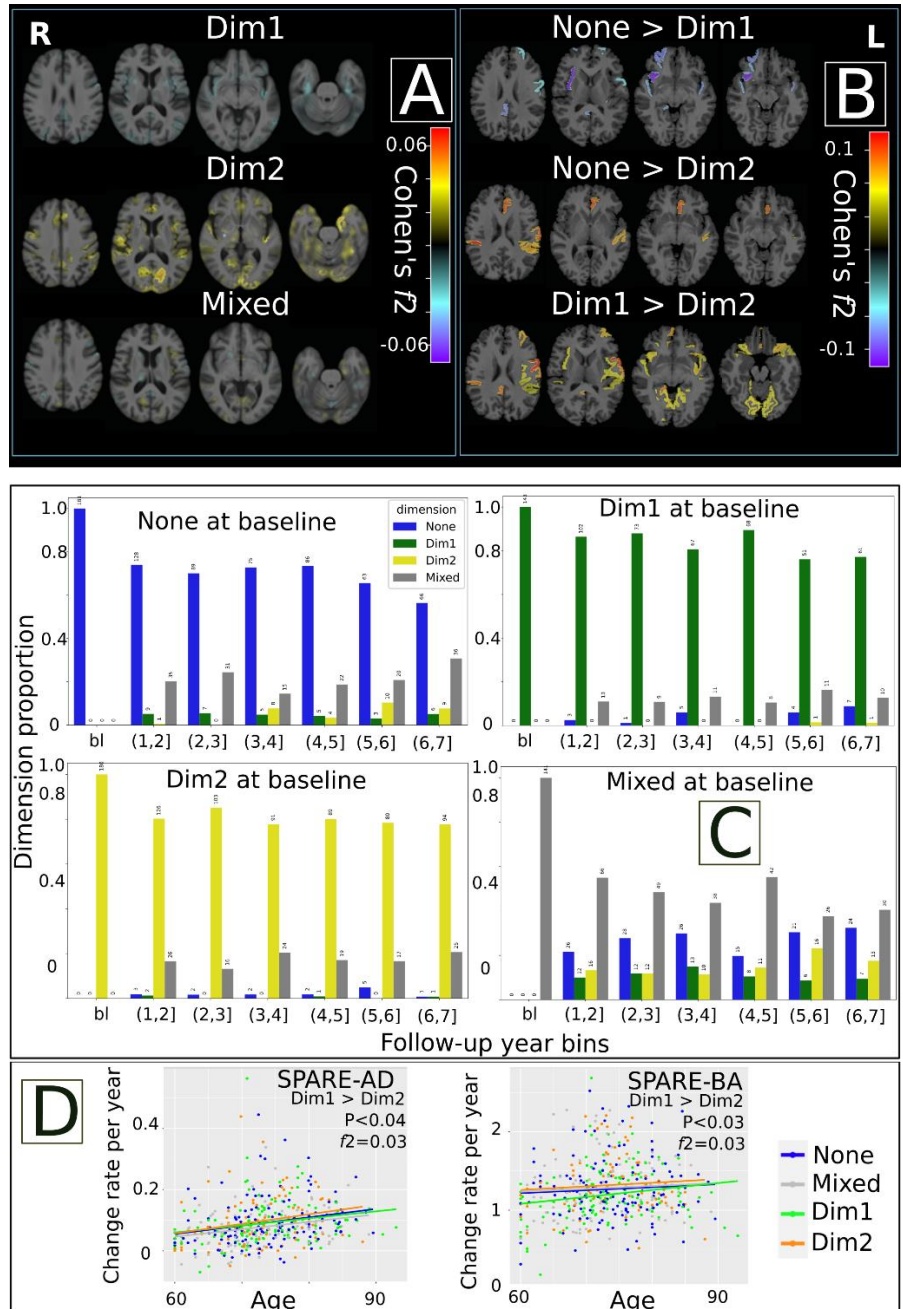

**Figure 4**: **A**) The two neuroanatomical dimensions in ADNI, BLSA, and BIOCARD baseline images show distinct grey matter abnormalities. Warmer color denotes brain atrophy (i.e., None > Dim), and cooler color represents larger tissue volume (i.e., Dim > None). Both directions are shown for each dimension. Cohen's $f2$ of ≥ 0.02, ≥ 0.15, and ≥ 0.35 signify small, moderate, and large effect sizes, respectively. L: left; R: right. **B**) The rate of change (RC) shows that Dim1's brain volume decreases with time more rapidly than Dim2. Only subjects for which MRI data were available at least for 6-time points were included for this analysis. **C**) Applying the HYDRA model to all available longitudinal scans with at least 6-7 years follow-ups. The two dimensions stay stable over time and are independent of each other. **D**) The positive RC for SPARE-AD and SPARE-BA of Dim2 is bigger than Dim1, meaning that Dim2 is more vulnerable to AD and brain aging longitudinally. Only subjects that have at least 6 time points were included for this analysis.

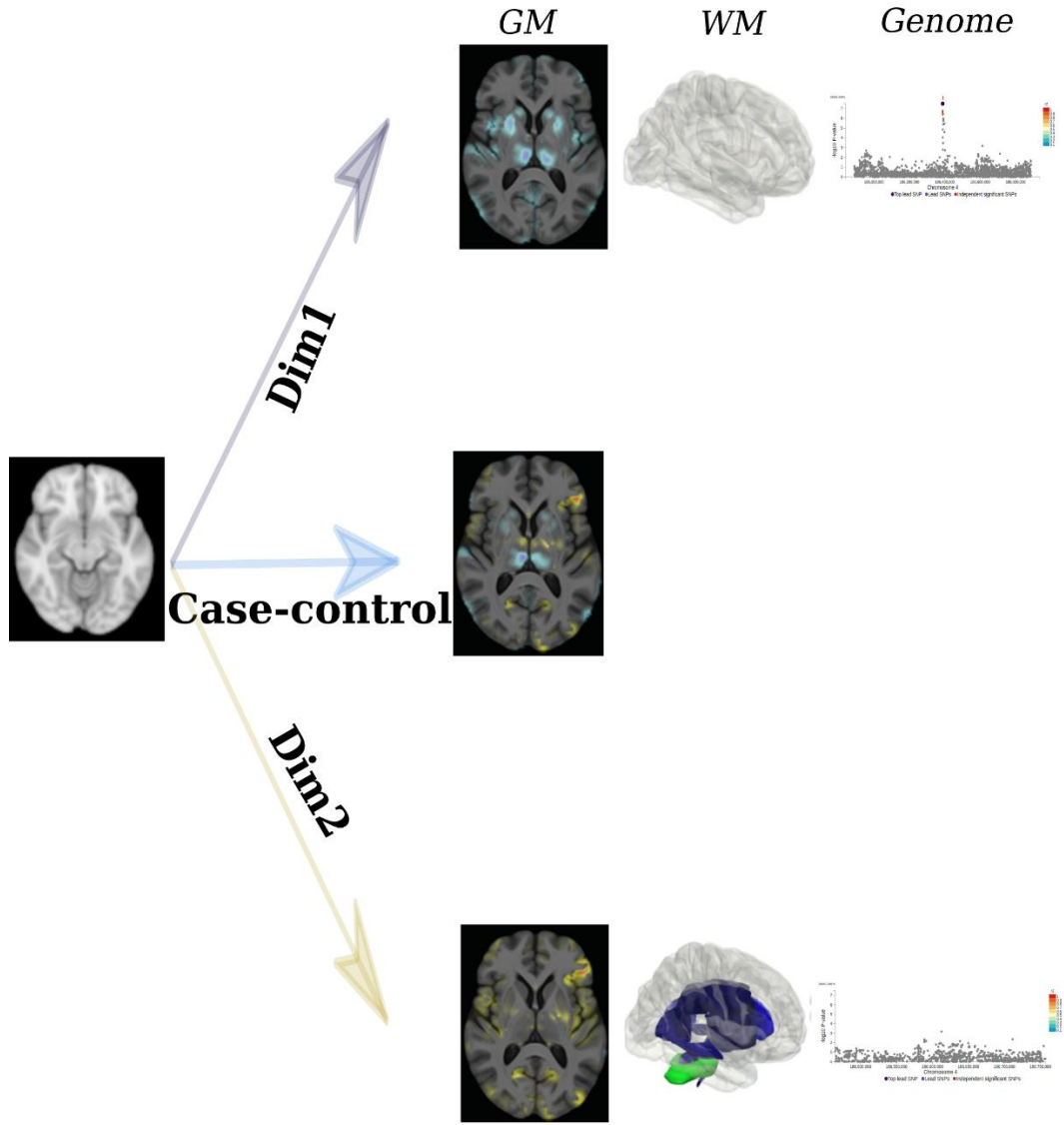

**Figure 5**: A two-axis framework to summarize the two-dimensional representation in LLD. The dimensional representation was anchored on the neuroanatomical heterogeneity in LLD. Normal brain anatomy is shown at the origin of the framework. We dissect case-control-based brain patterns (blue arrow) into two distinct dimensions, Dim1 (grey arrow) and Dim2 (yellow arrow). We then externally validated the two dimensions concerning microstructural WM integrity disruptions, where only Dim2 demonstrated this abnormality. Moreover, Dim2 was more affected by cognitive dysfunctionality and worse depressive severity. Lastly, for the genetic architectures, a *de novo* independent variant was significantly associated with Dim1 only, but not with Dim2.